\documentclass[aps,prd,twocolumn,showpacs,superscriptaddress]{revtex4-1}

\usepackage{graphicx}
\usepackage{amsmath,amssymb,bm}
\usepackage{multirow}
\usepackage{float}
\usepackage{hyperref}
\usepackage{booktabs}
\hypersetup{colorlinks=true,linkcolor=blue,citecolor=blue,urlcolor=blue}
\newcommand{\orcid}[1]{%
  \href{https://orcid.org/#1}{%
    \includegraphics[height=2ex, keepaspectratio]{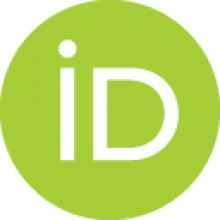}%
  }%
}

\newcommand{\R}{\mathbb{R}}

\begin{document}

\title{Contrastive self-supervised convolutional autoencoder for core-collapse supernova gravitational-wave detection}

\author{Tian-Yang Sun\orcid{0009-0002-5109-6420}}
%\email{suntianyang@stumail.neu.edu.cn}
\affiliation{Liaoning Key Laboratory of Cosmology and Astrophysics \& College of Sciences, Northeastern University, Shenyang 110819, China}

\author{Yue Niu\orcid{0009-0008-2974-8174}}
%\email{niuyue@mails.neu.edu.cn}
\affiliation{Liaoning Key Laboratory of Cosmology and Astrophysics \& College of Sciences, Northeastern University, Shenyang 110819, China}

\author{Chun-Yan Jiang\orcid{0009-0004-5008-1459}}
%\email{2450425551@qq.com}
\affiliation{Liaoning Key Laboratory of Cosmology and Astrophysics \& College of Sciences, Northeastern University, Shenyang 110819, China}

\author{Shang-Jie Jin\orcid{0000-0003-3697-3501}}
%\email{jinshangjie@stumail.neu.edu.cn}}
\affiliation{Liaoning Key Laboratory of Cosmology and Astrophysics \& College of Sciences, Northeastern University, Shenyang 110819, China}

\author{Yong Yuan\orcid{0009-0002-9540-9230}}
%\email{yuanyong@imech.ac.cn}
\affiliation{Center for Gravitational Wave Experiment, National Microgravity Laboratory, Institute of Mechanics, Chinese Academy of Sciences, Beijing, China}

%\author{Jing-Fei Zhang\orcid{0000-0002-3512-2804}}
%\email{jfzhang@mail.neu.edu.cn}
%\affiliation{Liaoning Key Laboratory of Cosmology and Astrophysics \& College of Sciences, Northeastern University, Shenyang 110819, China}

\author{Xin Zhang\orcid{0000-0002-6029-1933}}
\thanks{Corresponding author}
\email{zhangxin@neu.edu.cn}
\affiliation{Liaoning Key Laboratory of Cosmology and Astrophysics \& College of Sciences, Northeastern University, Shenyang 110819, China}
\affiliation{Key Laboratory of Data Analytics and Optimization for Smart Industry (Ministry of Education), Northeastern University, Shenyang 110819, China}
\affiliation{National Frontiers Science Center for Industrial Intelligence and Systems Optimization, Northeastern University, Shenyang 110819, China}

\begin{abstract}
Gravitational-wave astronomy has opened a direct observational window onto compact-object dynamics, strong-field gravity, and cosmology. Among the transient sources accessible through this window, core-collapse supernovae (CCSNe) are uniquely valuable because their signals can probe the engine of stellar collapse, proto-neutron-star dynamics, and explosion asymmetries, yet their weak, stochastic, and model-dependent waveforms remain difficult to detect. In this work, we develop a contrastive self-supervised convolutional autoencoder (CS-CAE) for CCSNe gravitational-wave signal detection. The method combines a convolutional autoencoder (CAE), a noise-centered latent regularizer, and a projection head trained with a contrastive objective. This design encourages independent noisy realizations of the same CCSNe signal to be mapped to nearby latent representations, thereby reducing the influence of random noise fluctuations. CS-CAE achieves performance comparable to a supervised convolutional neural network while clearly outperforming a conventional CAE baseline, and generalizes better to unseen numerical CCSNe waveform families. Under the Einstein Telescope (ET) detector configuration, the method achieves an effective sensitive distance of approximately $120\,\mathrm{kpc}$ and shows improved separation of CCSNe signals from stationary noise and transient glitches in the low-false-alarm regime. These results highlight the potential of CS-CAE as a robust and less template-dependent framework for CCSNe gravitational-wave searches.
\end{abstract}
\maketitle

\section{Introduction}
\label{sec:introduction}

The direct detection of gravitational waves (GWs) has opened a new observational window onto the dynamical Universe, enabling tests of general relativity in the strong-field regime~\cite{LIGOScientific:2016aoc,LIGOScientific:2016lio,Ezquiaga:2017ekz,LIGOScientific:2019fpa,Gong:2021jgg,LIGOScientific:2021sio}, independent constraints on compact-object populations, and new probes of cosmology and large-scale structure~\cite{LIGOScientific:2017adf,DES:2019ccw,KAGRA:2021duu}. In particular, GW standard sirens provide calibration-independent distance measurements and can be combined with other probes to break cosmological parameter degeneracies and constrain the late-time expansion history~\cite{Holz:2005df,Taylor:2012db,Chen:2017rfc,Wang:2019tto,Gray:2019ksv,You:2020wju,Yang:2021qge,Jin:2021pcv,Jin:2022qnj,Gair:2022zsa,Song:2022siz,Jin:2023sfc,Gray:2023wgj,Zhu:2023gmx,Jin:2025dvf}. These GW-based applications are closely connected to broader efforts to constrain dark energy, neutrino physics, and other late-Universe probes~\cite{Ma:2007av,WMAP:2008lyn,Cui:2009ns,Fu:2011ab,Lesgourgues:2012uu,Wyman:2013lza,Zhang:2014ifa,Lesgourgues:2014zoa,Planck:2015fie,Feng:2017nss,Vagnozzi:2017ovm,Feng:2017usu,Planck:2018vyg,Guo:2018gyo,Vagnozzi:2018jhn,H0LiCOW:2019pvv,Feng:2019jqa,Macquart:2020lln,Zhao:2020ole,Wang:2021kxc,Li:2024qso,Li:2024qus,DESI:2024mwx,Li:2025owk,DESI:2025zgx,Li:2026xaz}. The current ground-based detector network, including LIGO~\cite{LIGOScientific:2014pky}, Virgo~\cite{VIRGO:2014yos}, and KAGRA~\cite{KAGRA:2020tym}, has firmly established the detection of compact-binary coalescences and has progressively improved its sensitivity across successive observing runs~\cite{LIGOScientific:2025slb}. Beyond these realized detections, next-generation ground-based observatories, such as ET~\cite{Punturo_2010,Hild:2010id} and Cosmic Explorer (CE)~\cite{Reitze:2019iox,Evans:2021gyd}, are expected to substantially enlarge the accessible source volume and improve the prospects for detecting weak or rare transient sources~\cite{Srivastava:2019fcb,ET:2025xjr}. This expected expansion of the observable transient population motivates detection methods that are not limited to well-modeled, phase-coherent chirp signals~\cite{Drago:2020kic,KAGRA:2021tnv,KAGRA:2025yfg}. Core-collapse supernovae (CCSNe) are a particularly important target in this context, because their GW emission may be accompanied by electromagnetic and neutrino counterparts and can provide a direct multimessenger view of stellar death~\cite{Gossan:2015xda,Halim:2021yqa,Szczepanczyk:2023ihe}.

CCSNe arise from the gravitational collapse of massive stellar cores and involve a sequence of strongly coupled physical processes, including core bounce, shock stalling and revival, turbulent convection, standing accretion shock instability (SASI), proto-neutron-star (PNS) oscillations, rotation, magnetic fields, and anisotropic neutrino emission~\cite{Abdikamalov:2020jzn}. Their GW signals can encode information that is largely inaccessible through electromagnetic observations alone~\cite{Vartanyan:2023sxm}. Direct detection of CCSNe GWs would therefore constrain stellar evolution, nuclear microphysics, nucleosynthesis conditions, neutron-star formation, and black-hole formation in a way that is complementary to optical, neutrino, and high-energy observations~\cite{Logue:2012zw,Powell:2025oig}. The challenge is that the predicted CCSNe GW strain is weak and its morphology depends sensitively on progenitor mass, equation of state, rotation, numerical transport scheme, dimensionality, and viewing angle~\cite{Abdikamalov:2020jzn}.

Searches for CCSNe GWs have traditionally relied on minimally modeled burst strategies, including excess-power statistics, coherent network likelihoods, time-frequency clustering, and Bayesian wavelet reconstruction~\cite{Anderson:2000yy,Klimenko:2008fu,Sutton:2009gi,Heng:2009zz,Rover:2009ia,Thrane:2013bea,Cornish:2014kda,Klimenko:2015ypf,Millhouse:2018dgi,Roma:2019kcd,Suvorova:2019ebd,Drago:2020kic,Takeda:2021hmf,Raza:2022kcs,Yuan:2023umh,Pastor-Marcos:2023tcc,Yuan:2024bxq,Veutro:2025mrv}. These methods underpin both all-sky short-duration burst searches and externally triggered searches for optically observed CCSNe in LIGO-Virgo-KAGRA data~\cite{LIGOScientific:2016jvu,LIGOScientific:2019ryq,KAGRA:2021tnv,Szczepanczyk:2023ihe,KAGRA:2025yfg}. Matched filtering is optimal under stationary Gaussian noise when accurate templates are available~\cite{Allen:2005fk}, and recent work has explored theoretically informed template banks for CCSNe GW searches~\cite{Andresen:2024wrc}, but it remains challenging for CCSNe because the relevant waveform family is broad, stochastic, and incompletely known. Even if phenomenological templates are constructed to capture PNS frequency tracks or other dominant components, these templates cannot fully represent the broad waveform diversity arising from the strongly coupled physical processes involved in CCSNe~\cite{Cerda-Duran:2025new}. Unmodeled burst searches are more model-agnostic, but they may lose sensitivity in the low signal-to-noise-ratio (SNR) regime and can be affected by non-Gaussian transient artifacts, or glitches, that can resemble short astrophysical bursts~\cite{Zevin:2016qwy,Cabero:2019orq,Szczepanczyk:2021bka}. These considerations have motivated the use of deep learning for fast detection statistics, representation learning, and classification in CCSNe GW data analysis~\cite{Astone:2018uge,Chan:2019fuz,Iess:2020yqj,LopezPortilla:2020odz,Cavaglia:2020qzp,Edwards:2020hmd,Saiz-Perez:2021bce,Antelis:2021qak,Chao:2022tui,Mitra:2022nyc,Lagos:2023qli,Mitra:2023ylb,Iess:2023quq,Powell:2023bex,Morales:2024cme,Nunes:2024fig,Abylkairov:2024hjf,Veutro:2025fze,Wang:2026iyy,Akhmetali:2026yns,Akhmetali:2026hbf}.

Deep learning has been explored broadly in GW, astronomical, and cosmological data analysis~\cite{Razzano:2018fxb,Krastev:2019koe,Cuoco:2020ogp,Green:2020dnx,Ormiston:2020ele,Yu:2021swq,Dax:2021tsq,Zhao:2022qob,Dax:2022pxd,Saleem:2023hcm,Wang:2023lif,Sun:2023vlq,Xiong:2024gpx,Dax:2024mcn,Wang:2024oei,Sun:2024ywb,Cuoco:2024cdk,Qin:2025mvj,Sun:2025ypd,Zhang:2026okw,Liu:2026nhu,Sun:2026dga}. Many GW applications use supervised models, such as convolutional neural networks (CNN) trained on labeled signal and noise samples~\cite{Gabbard:2017lja,George:2017pmj,Chan:2019fuz,Iess:2020yqj,LopezPortilla:2020odz}. These models can learn nonlinear features from time-domain strain data or time-frequency representations and can provide fast detection statistics once trained~\cite{Iess:2023quq}. However, their performance is tied to the labeled waveform bank and detector-noise model used during training~\cite{Gebhard:2019ldz,Chan:2019fuz}. This limitation is particularly important for CCSNe, because reliable labels are obtained mainly from simulations and the predicted waveforms depend on progenitor structure, equation of state, rotation, dimensionality, neutrino transport, and viewing angle~\cite{Ott:2008wt,Kotake:2011yv,Richers:2017joj,Andresen:2018aom}. A supervised classifier that performs well on one waveform family may therefore generalize poorly to waveforms generated under different physical or numerical assumptions.

Autoencoder-based self-supervised learning offers a natural way to reduce the dependence of GW searches on labeled signal banks and precomputed templates~\cite{Morawski:2021kxv,Moreno:2021fvp}. A recent autoencoder-based framework shows that learned low-dimensional representations can provide useful separation of compact-binary signals, detector glitches, and unmodeled transients~\cite{Raikman:2023ktu}. In parallel, contrastive self-supervised learning has been explored for GW signal identification by associating different noisy or augmented views of the same underlying signal~\cite{Liu:2023oau}. However, contrastive latent representation learning for CCSNe GW detection remains less explored. This is an important gap because CCSNe GW signals are stochastic, and strongly dependent on uncertain source physics, requiring representations that are robust to detector noise and transferable across waveform morphologies.

In this work, we develop a contrastive self-supervised convolutional autoencoder (CS-CAE) for CCSNe GW detection. The model extends a reconstruction-based convolutional autoencoder (CAE) by combining a reconstruction loss, a noise-centered latent regularization term, and an InfoNCE-style contrastive objective. The contrastive pairs are constructed from independent detector-noise realizations of the same underlying CCSNe waveform, so that the encoder is encouraged to learn signal-consistent representations that are less sensitive to incidental noise fluctuations. We compare CS-CAE with a supervised convolutional neural network and a CAE baseline on in-distribution simulated data, cross-distribution numerical CCSNe waveforms, and glitch-contaminated test samples. In particular, we assess whether contrastive self-supervised representation learning can improve generalization across waveform families and robustness to glitches while maintaining competitive detection sensitivity.

\section{Methods}
\label{sec:methods}

\subsection{Data simulation and preprocessing}
\label{subsec:data_simulation}

We use a phenomenological CCSNe source model and an ET detector-noise model to construct the training data. The overall data-generation procedure follows the standard gravitational-wave injection pipeline: source-frame polarizations are projected into the detector channels using antenna response functions, and the resulting detector-frame strains are whitened using the target one-sided noise power spectral density. Each detector-frame sample is represented as a three-channel time series corresponding to the ET interferometer channels $\{\mathrm{E1}, \mathrm{E2}, \mathrm{E3}\}$~\cite{Hild:2010id,ET:2025xjr}.

The source-frame waveform contains two polarizations, $h_+(t)$ and $h_\times(t)$. For the $k$-th ET channel, the projected detector-frame strain can be written schematically as
\begin{equation}
    h_k(t)
    =
    F_{+,k}(\alpha,\delta,\psi,t_{\rm gps}) h_+(t)
    +
    F_{\times,k}(\alpha,\delta,\psi,t_{\rm gps}) h_\times(t),
    \label{eq:detector_projection}
\end{equation}
where $F_{+,k}$ and $F_{\times,k}$ are the detector antenna pattern functions, and $\alpha$, $\delta$, $\psi$, and $t_{\rm gps}$ denote the right ascension, declination, polarization angle, and GPS time, respectively~\cite{Jaranowski:1998qm}. During data generation, the sky location is sampled isotropically and the GPS time is randomized to vary the detector response. The detector response is computed using the \texttt{PyCBC} detector module~\cite{alex_nitz_2024_10473621}. The waveform amplitude is scaled according to the inverse-distance dependence of gravitational-wave strain. When the waveform is expressed in the commonly used distance-scaled $hD$ normalization, the detector-frame strain is obtained by dividing by the source distance~\cite{Gossan:2015xda,Vartanyan:2023sxm,Cerda-Duran:2025new}.

The phenomenological CCSNe source model used for the training set is designed to reproduce characteristic time-frequency features reported in multidimensional CCSNe simulations, rather than to perform a full radiation-hydrodynamics simulation~\cite{Murphy:2009dx,Mueller:2012sv,Morozova:2018glm,Radice:2018usf,Mezzacappa:2022xmf,Vartanyan:2023sxm,Murphy:2025vyv}. In particular, the model includes components motivated by the rising high-frequency emission associated with proto-neutron-star (PNS) oscillations, low-frequency SASI/convection activity, rapid-rotation bounce-like emission, and broadband stochastic emission. The resulting waveform is written as
\begin{equation}
    h_{+,\times}(t)
    =
    h^{\rm PNS}_{+,\times}(t)
    +
    h^{\rm SASI}_{+,\times}(t)
    +
    h^{\rm rot}_{+,\times}(t)
    +
    h^{\rm stoch}_{+,\times}(t),
    \label{eq:CCSNe_components}
\end{equation}
where the PNS component is parameterized by a time-dependent central frequency,
\begin{equation}
    f_{\rm PNS}(t)
    =
    f_0+\dot f t+\frac{1}{2}\ddot f t^2 .
    \label{eq:pns_frequency}
\end{equation}
This parameterization is used as a simplified representation of the upward-drifting high-frequency emission commonly observed in CCSNe simulations. Stochastic components are added around this central frequency track to emulate the broad and non-phase-coherent morphology of CCSNe GW emission. The simulated population therefore includes PNS-dominated stochastic emission, broadband high-frequency components, SASI-like low-frequency emission when present, and possible rapid-rotation bounce-like bursts~\cite{Hayama:2016kmv,Chan:2020gnx,Takeda:2021hmf,Murphy:2025vyv,Veutro:2025mrv}. In the default training configuration, the memory-like component is not included.

Detector noise is generated as Gaussian noise consistent with the ET-D amplitude spectral density and independently sampled for the three ET channels. The data are whitened in the frequency domain using the corresponding one-sided noise power spectral density, following standard GW data-analysis practice~\cite{Hild:2010id,Usman:2015kfa,Biwer:2018osg}. The frequency band used for noise generation, whitening, and SNR evaluation is $10$--$2048\,\mathrm{Hz}$. For a time series $d(t)$ with Fourier transform $\tilde d(f)$, the whitened data are computed as
\begin{equation}
    \tilde d_{\rm w}(f)
    =
    \frac{\tilde d(f)}
    {\sqrt{\frac{1}{2} f_s S_n(f)}} ,
    \label{eq:whitening}
\end{equation}
where $f_s$ is the sampling rate and $S_n(f)$ is the one-sided noise power spectral density.

\subsection{Deep-learning model architectures}
\label{subsec:architectures}

\begin{figure*}[t]
    \centering  \includegraphics[width=0.8\textwidth]{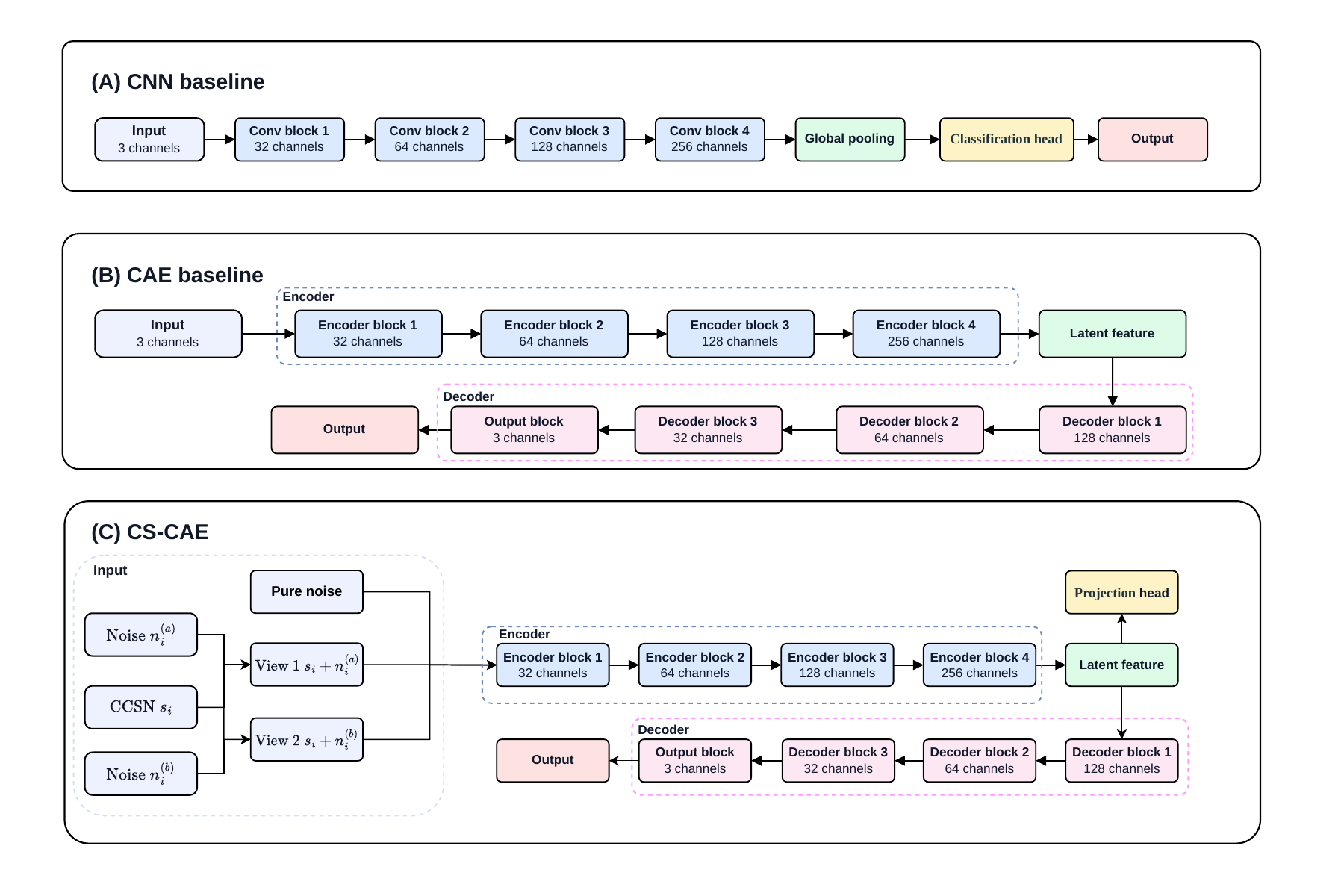}
    \caption{Network architectures of the supervised CNN baseline, the CAE baseline, and the proposed CS-CAE. Panel (A) shows the CNN classifier, panel (B) shows the CAE baseline, and panel (C) shows the CS-CAE. Light blue blocks denote inputs, blue blocks denote convolutional or encoder blocks, green blocks denote pooled or latent features, yellow blocks denote classification or projection heads, pink blocks denote decoder blocks, and light red blocks denote outputs. In panel (C), the dashed box indicates the construction of two noisy views from the same CCSNe waveform and independent noise realizations.}
    \label{fig:net}
\end{figure*}

We compare three neural-network architectures, as summarized in Fig.~\ref{fig:net}: a supervised CNN classifier, a reconstruction-based CAE baseline, and the proposed CS-CAE model. The three models share the same four-stage one-dimensional convolutional backbone for a fair comparison, but differ in their output heads, detection statistics, and training objectives.

\subsubsection{Convolutional neural network baseline}
\label{subsubsec:supervised_cnn}

The first baseline is a conventional supervised CNN classifier, shown in Fig.~\ref{fig:net}(A). The input is a three-channel whitened time series $x\in\R^{3\times N}$, and the network outputs two logits for the noise and signal classes. The convolutional backbone consists of four one-dimensional convolutional blocks with channel widths $32$, $64$, $128$, and $256$, kernel sizes $16$, $8$, $8$, and $4$, a stride of $2$ in each block, and padding values of $7$, $3$, $3$, and $1$, respectively. Each block contains convolution, batch normalization, and ReLU activation, and the first three blocks are followed by max pooling with kernel size $2$. An adaptive average pooling layer maps the final feature map to a fixed-length vector, and the classification head consists of a linear layer with $128$ hidden units, a ReLU activation, dropout with rate $0.2$, and a final linear layer producing two class logits. The signal probability is obtained by applying a softmax function,
\begin{equation}
\hat y_i
=
P(y_i=1\mid x_i)
=
\frac{\exp(\ell_{i,1})}
{\sum_{c=0}^{1}\exp(\ell_{i,c})},
\quad
\ell_i=g_{\rm sup}(E_\theta(x_i)),
\label{eq:supervised_output}
\end{equation}
where $E_\theta$ denotes the CNN backbone, $g_{\rm sup}$ is the supervised classification head, and $\ell_{i,1}$ denotes the logit of the signal class. The supervised baseline is trained with the cross-entropy loss,
\begin{equation}
    \mathcal{L}_{\rm CE}
    =
    -\frac{1}{B}\sum_{i=1}^{B}
    \log
    \frac{\exp(\ell_{i,y_i})}
    {\sum_{c=0}^{1}\exp(\ell_{i,c})},
    \label{eq:ce}
\end{equation}
where $\ell_{i,c}$ is the logit of class $c$ for the $i$-th sample.

\subsubsection{Convolutional autoencoder baseline}
\label{subsubsec:ordinary_ssl}

The second baseline is a reconstruction-based CAE, shown in Fig.~\ref{fig:net}(B). Its encoder uses the same four convolutional channel widths, kernel sizes, strides, and padding values as the CNN backbone, but does not include the max-pooling layers used in the supervised CNN classifier. After adaptive average pooling, a linear embedding layer maps the encoder output to a $128$-dimensional normalized latent feature. The decoder upsamples the encoded feature map through four linear upsampling stages with scale factor $2$, followed by one-dimensional convolutional layers with kernel size $7$ and padding $3$. The first three decoder blocks use channel widths $128$, $64$, and $32$ with batch normalization and ReLU activation, and the final output block maps the feature map back to the original three channels. Its self-supervised objective is to reconstruct pure-noise inputs,
\begin{equation}
    \hat x_i = D_\phi(E_\theta(x_i)),
    \label{eq:ae_reconstruction}
\end{equation}
with mean-squared reconstruction loss
\begin{equation}
    \mathcal{L}_{\rm rec}
    = \frac{1}{B}\sum_{i=1}^{B}\left\|x_i-\hat x_i\right\|_2^2 .
    \label{eq:rec_loss}
\end{equation}

After training, the CAE is used directly as a reconstruction-based detector rather than being followed by an additional classification head. The detection statistic is the mean-squared reconstruction error,
\begin{equation}
    S_{\rm CAE}(x)
    =
    \frac{1}{3N}\left\|x-D_\phi(E_\theta(x))\right\|_2^2 .
\label{eq:cae_score}
\end{equation}
The CAE baseline therefore isolates the behavior of a purely reconstruction-based self-supervised detector without the contrastive term introduced in the proposed model.

\subsubsection{Contrastive self-supervised convolutional autoencoder}
\label{subsubsec:proposed_model}

The proposed CS-CAE, shown in Fig.~\ref{fig:net}(C), uses the same encoder--decoder structure as the CAE baseline and further adds a projection head for contrastive learning. The projection head is a two-layer multilayer perceptron consisting of a linear layer from $128$ to $128$ dimensions, a ReLU activation, and a final linear layer from $128$ to $128$ dimensions. The encoder first maps an input time series to a convolutional feature map and then to a $128$-dimensional normalized embedding,
\begin{equation}
    z_i=\frac{W_{\rm emb}\,\mathrm{Pool}(E_\theta(x_i))}{\left\|W_{\rm emb}\,\mathrm{Pool}(E_\theta(x_i))\right\|_2} .
    \label{eq:embedding}
\end{equation}
The projection head $q_\psi$ maps $z_i$ to a normalized contrastive vector,
\begin{equation}
    p_i=\frac{q_\psi(z_i)}{\left\|q_\psi(z_i)\right\|_2} .
    \label{eq:projection}
\end{equation}

For contrastive pretraining, the projected signal waveform $s_i$ is combined with two independently sampled pure-noise realizations, $n_i^{(a)}$ and $n_i^{(b)}$, to construct two noisy views,
\begin{equation}
    x_i^{(1)} = s_i+n_i^{(a)}, \qquad
    x_i^{(2)} = s_i+n_i^{(b)}.
    \label{eq:positive_views}
\end{equation}
These two views form a positive pair because they share the same underlying CCSNe waveform while differing in detector-noise realization. This construction follows the general contrastive-learning principle of bringing the representations of positive pairs closer together in representation space, but adapts the positive-pair definition to CCSNe signal injections in independent detector-noise realizations~\cite{DBLP:journals/corr/abs-1807-03748,DBLP:conf/icml/ChenK0H20,Liu:2023oau}.

For a mini-batch containing $B$ clean signals and two independent noise realizations per signal, the positive pair is $(p_i^{(1)},p_i^{(2)})$, and the remaining $2B-2$ views act as negatives. The contrastive loss is an InfoNCE-style cross-entropy loss~\cite{DBLP:journals/corr/abs-1807-03748},
\begin{equation}
    \mathcal{L}_{\rm con}
    = -\frac{1}{2B}\sum_{i=1}^{2B}
    \log
    \frac{\exp\left(\mathrm{sim}(p_i,p_{i^+})/\tau\right)}
    {\sum_{j=1,\,j\neq i}^{2B}\exp\left(\mathrm{sim}(p_i,p_j)/\tau\right)} ,
    \label{eq:contrastive_loss}
\end{equation}
where $i^+$ is the paired view of the same signal, $\mathrm{sim}(a,b)=a^\mathsf{T}b$ for normalized vectors, and $\tau=0.1$.

A noise-center regularizer is also used to concentrate pure-noise embeddings around an exponential-moving-average noise center $c_n$~\cite{DBLP:conf/icml/RuffGDSVBMK18},
\begin{equation}
    \mathcal{L}_{\rm cen}
    = \frac{1}{B}\sum_{i=1}^{B}\left\|z_i^{\rm noise}-c_n\right\|_2^2 .
    \label{eq:center_loss}
\end{equation}

In each mini-batch, the two pure-noise realizations are concatenated and used to compute the reconstruction loss and the noise-center regularization. The same clean signal combined with the two independent noise realizations forms a positive pair for the contrastive loss. The total training objective is
\begin{equation}
\mathcal{L}
=
\lambda_{\rm rec}\mathcal{L}_{\rm rec}
+
\lambda_{\rm cen}\mathcal{L}_{\rm cen}
+
\lambda_{\rm con}\mathcal{L}_{\rm con}.
\label{eq:total_loss}
\end{equation}

In this work, $\lambda_{\rm rec}=1.0$, $\lambda_{\rm cen}=0.1$, and $\lambda_{\rm con}=1.0$, according to the overall validation performance. After training, CS-CAE is used without an additional classification head. Its detection statistic combines the distance to the learned noise center and the reconstruction error,
\begin{equation}
S_{\rm CS\text{-}CAE}(x)
=
\left\|z(x)-c_n\right\|_2
+
\beta_{\rm rec}
\frac{1}{3N}\left\|x-D_\phi(E_\theta(x))\right\|_2^2 ,
\label{eq:cscae_score}
\end{equation}
with $\beta_{\rm rec}=0.5$.

\subsection{Experimental setup}
\label{subsec:experimental_setup}

At each training epoch, $10^4$ projected CCSNe signal samples and two independent sets of $10^4$ pure-noise samples are generated. For the supervised CNN, the first noise set is added to the CCSNe signals to construct $10^4$ noisy signal samples, while the second noise set provides $10^4$ pure-noise samples. For the CAE and CS-CAE, the two noise sets serve as independent noise realizations, and noisy CCSNe views are constructed during training as needed. The sampling rate is $4096~\mathrm{Hz}$, the full detector-frame data segment has a duration of $1.5~\mathrm{s}$, and the intrinsic CCSNe waveform duration is $1\,\mathrm{s}$. The injection start time is randomized in the interval $[0.2, 0.3]\,\mathrm{s}$. The source distance is sampled from a log-uniform prior over $[1, 200]\,\mathrm{kpc}$, and signal samples with network SNR below $8$ are rejected. For the test sets containing CCSNe injections, signal samples with network SNR below $5$ are rejected. In the training data generated at each epoch, the generated signal set consists of whitened projected CCSNe waveforms before noise addition, while the two noise sets consist of independent pure detector-noise realizations.

To evaluate generalization beyond the phenomenological waveform distribution used for training, we construct a cross-distribution test set from independent numerical CCSNe waveform catalogs. Specifically, we use the waveform families summarized in Table~I of Ref.~\cite{Wang:2026iyy}. The selected catalogs include rotational and neutrino-driven CCSNe waveforms from multiple simulation studies~\cite{Muller:2011yi,Ott:2012mr,Kuroda:2017trn,Powell:2018isq,Andresen:2018aom,Radice:2018usf,Mezzacappa:2020lsn,Powell:2020cpg,Vartanyan:2023sxm}\footnote{\protect\url{https://wwwmpa.mpa-garching.mpg.de/ccsnarchive/data/Andresen2019/}, \protect\url{https://www.astro.princeton.edu/~burrows/gw.3d/}}. These catalogs cover progenitor masses from $9~M_\odot$ to $60~M_\odot$ and contain waveform morphologies that are not generated by the phenomenological source model used for training.

We also construct a glitch-contaminated test set as a generic stress test of robustness against non-Gaussian transient noise. The glitch waveforms are simulated with \texttt{gengli}, a generative package designed to produce blip-like transient noise artifacts in gravitational-wave detectors~\cite{Lopez:2022lkd,Lopez:2022dho}. The glitch injections have SNRs in the range $8$--$30$, with parameters $\alpha=0.2$ and $f_{\rm high}=250.0\,\mathrm{Hz}$. The glitch-contaminated test set contains pure-noise samples and noise-plus-glitch samples. The noise-plus-glitch samples are used to quantify the rate at which transient instrumental artifacts are misidentified as CCSNe signals and to examine whether the learned latent representations separate CCSNe waveforms from glitch-dominated backgrounds.

Table~\ref{tab:dataset_summary} summarizes the datasets used in this work. Here, Test I denotes the in-distribution test set, Test II denotes the cross-distribution numerical-waveform test set, and Test III denotes the glitch-robustness test set.

\begin{table*}[htbp]
\caption{Summary of the datasets used for training and evaluation.}\label{tab:dataset_summary}
\centering
\setlength\tabcolsep{14pt}
\renewcommand{\arraystretch}{1.5}
\begin{tabular}{lcc}
\hline \hline 
Dataset & Purpose & Composition \\
\hline
Training & Model training & Phenomenological CCSNe injections \\
Test I & In-distribution evaluation & Independent phenomenological CCSNe injections \\
Test II & Cross-distribution evaluation & Numerical CCSNe waveform injections from independent catalogs \\
Test III & Glitch robustness evaluation & \texttt{Gengli}-simulated noise-plus-glitch samples \\
\hline \hline
\end{tabular}
\end{table*}

All neural-network models are implemented in PyTorch~\cite{DBLP:conf/nips/PaszkeGMLBCKLGA19} and are trained and evaluated under the same preprocessing and metric definitions. The CNN is trained with the cross-entropy loss, while the CAE and CS-CAE are trained with the self-supervised objectives described above. The models are optimized with AdamW~\cite{DBLP:conf/iclr/LoshchilovH19} using a learning rate of $10^{-3}$, weight decay of $10^{-4}$, batch size of $512$, and $100$ training epochs. We use the PyTorch \texttt{ReduceLROnPlateau} learning-rate scheduler, monitoring the validation loss, with reduction factor $0.5$ and patience $5$. The training/validation split is $80\%/20\%$.

The detection performance is primarily evaluated using the receiver operating characteristic (ROC) curve and the corresponding area under the curve (AUC). For each model, a scalar detection statistic is assigned to every input sample. For the supervised CNN, this statistic is the signal probability, while for the CAE and CS-CAE it is the corresponding anomaly score defined above. By fixing a threshold $S^\ast$, samples with $S>S^\ast$ are classified as CCSNe signal candidates, whereas samples with $S<S^\ast$ are classified as noise.

From the entries of the confusion matrix, we define the detection efficiency, which is equivalent to the true positive rate (TPR), as
\begin{equation}
    \mathrm{TPR} =
    \frac{\mathrm{TP}}{\mathrm{TP}+\mathrm{FN}},
    \label{eq:efficiency}
\end{equation}
and the false positive rate (FPR) as
\begin{equation}
    \mathrm{FPR} =
    \frac{\mathrm{FP}}{\mathrm{TN}+\mathrm{FP}},
    \label{eq:fpr}
\end{equation}
where $\mathrm{TP}$, $\mathrm{FN}$, $\mathrm{FP}$, and $\mathrm{TN}$ denote true positives, false negatives, false positives, and true negatives, respectively. The ROC curve is then obtained by varying the threshold $S^\ast$ and plotting the detection efficiency $\mu$ against the false positive rate, FPR.

\section{Results}
\label{sec:results}

\subsection{Detection performance}
\label{subsec:overall_detection}

We first evaluate the detection performance of the three models using ROC curves under two complementary test scenarios. The first scenario adopts an in-distribution simulated test set generated according to the same data-generation prescription as the training samples. This setting evaluates the detection performance of each model on test data drawn from the same distribution as the training set. The second scenario employs numerical CCSNe waveforms that are not generated by the training simulation prescription, thereby providing a more stringent assessment of cross-waveform and cross-simulation generalization.

\begin{figure}[t]
    \centering
    \includegraphics[width=0.48\textwidth]{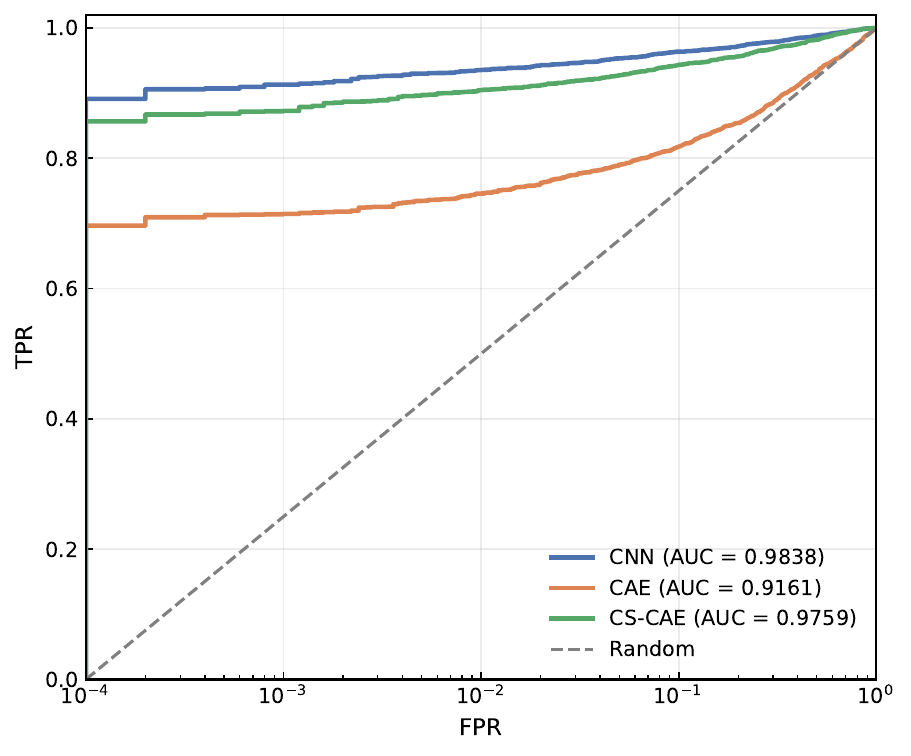}
    \caption{ROC curves on the in-distribution simulated test set generated with the same prescription as the training data. The blue, orange, and green solid curves correspond to the CNN, CAE, and CS-CAE, respectively. The gray dashed diagonal line denotes random classification, and the legend lists the AUC value for each model.}
    \label{fig:roc_test}
\end{figure}

Figure~\ref{fig:roc_test} presents the ROC curves obtained on the in-distribution simulated test set. In this matched setting, the supervised CNN achieves the highest area under the curve, with $\mathrm{AUC}=0.9838$, followed by the proposed CS-CAE with $\mathrm{AUC}=0.9759$ and the CAE with $\mathrm{AUC}=0.9161$. The superior performance of the supervised CNN is expected, because its discriminative decision boundary is directly optimized using binary labels drawn from the same simulation prescription as the test set. The more relevant observation is that CS-CAE substantially improves upon the conventional self-supervised CAE baseline and achieves performance comparable to that of the supervised CNN, despite not relying on the same fully supervised discriminative objective. This indicates that the contrastive learning constraint effectively enhances the signal-sensitive representation learned by the autoencoder, enabling the self-supervised framework to approach the detection capability of a label-trained CNN in the matched in-distribution setting.

\begin{figure}[t]
    \centering
    \includegraphics[width=0.48\textwidth]{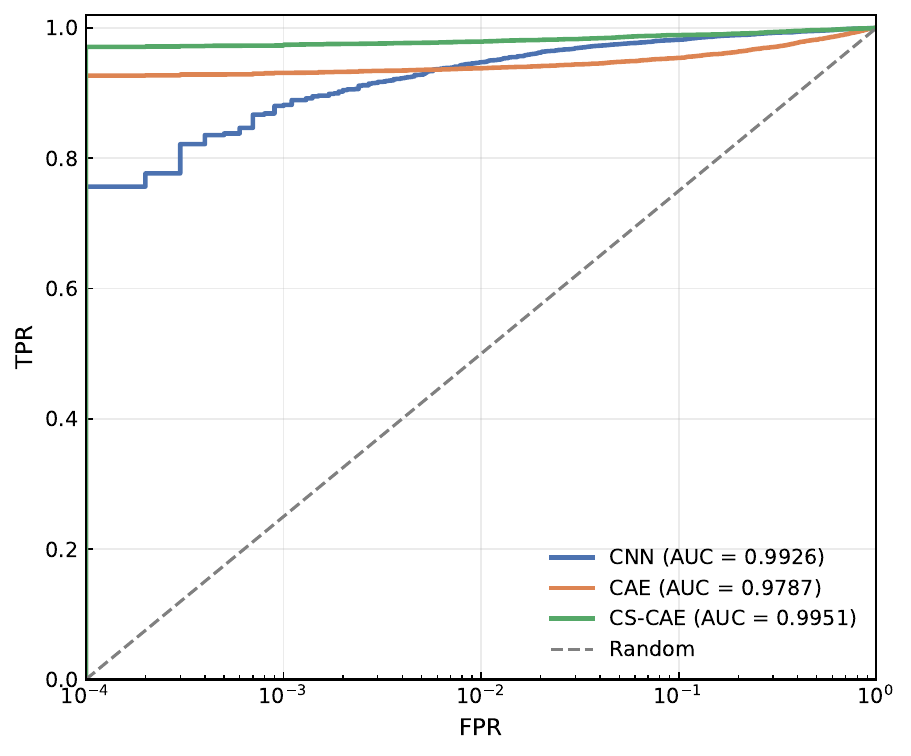}
    \caption{ROC curves on the numerical CCSNe waveform test set. The blue, orange, and green solid curves correspond to the CNN, CAE, and CS-CAE, respectively. The gray dashed diagonal line denotes random classification, and the legend lists the AUC value for each model.}
    \label{fig:roc_sim}
\end{figure}

A different performance hierarchy is observed when the models are evaluated on the numerical CCSNe waveform test set. As shown in Fig.~\ref{fig:roc_sim}, CS-CAE achieves the highest AUC, $\mathrm{AUC}=0.9951$, compared with $\mathrm{AUC}=0.9926$ for the supervised CNN and $\mathrm{AUC}=0.9787$ for the CAE. Although the absolute difference between CS-CAE and the CNN is modest, the reversal in ranking is noteworthy. It suggests that the representation learned through contrastive self-supervision transfers more effectively when the CCSNe waveform morphology differs from that of the phenomenological simulations used during training. In this sense, the supervised CNN performs slightly better under the in-distribution evaluation, whereas CS-CAE exhibits stronger robustness under the numerical-waveform evaluation.

This behavior can be understood from the nature of the contrastive objective. During training, CS-CAE constructs positive pairs from independent noise realizations of the same underlying waveform. This encourages the encoder to retain features that are invariant to detector-noise fluctuations rather than features specific to the phenomenological waveform morphology. As a result, the learned representations capture more general signal characteristics, such as multi-channel coherence and time-frequency structure, that are shared across different CCSNe waveform families. In contrast, the supervised CNN is trained with binary labels derived from the phenomenological simulation prescription, so its decision boundary is more tightly coupled to the morphological features of that specific waveform family. This distributional coupling explains why the supervised CNN achieves higher AUC on the in-distribution test set but is surpassed by CS-CAE on the numerical waveform test set.

Taken together with the in-distribution results, this comparison shows that the proposed CS-CAE does not merely optimize detection performance under a single matched distribution. Instead, it retains sensitivity comparable to the supervised CNN in the matched test setting while achieving improved transfer performance on the numerical CCSNe waveform set. This behavior is particularly relevant for CCSNe GW searches, where the true astrophysical waveform family is uncertain and cannot be exhaustively represented by a single simulation prescription.

\subsection{Distance-dependent sensitivity}
\label{subsec:distance_lowfar}

\begin{figure}[t]
    \centering
    \includegraphics[width=0.48\textwidth]{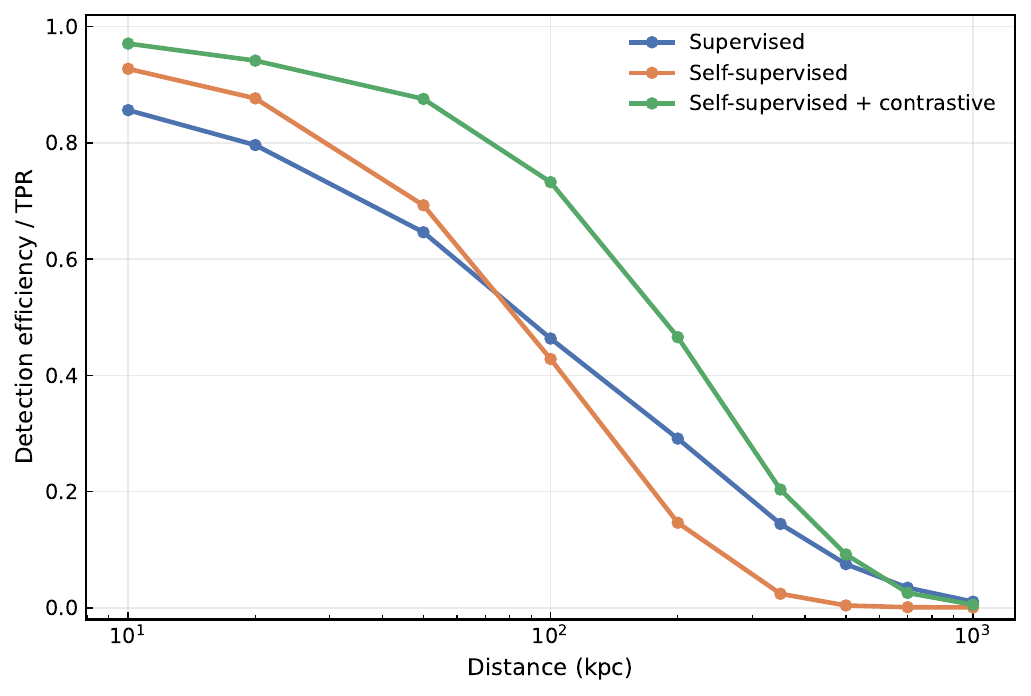}
    \caption{Detection efficiency as a function of source distance. The vertical axis gives the true positive rate at the adopted operating threshold, and the horizontal axis gives the source distance in $\mathrm{kpc}$ on a logarithmic scale. The blue, orange, and green curves correspond to the supervised CNN, the self-supervised CAE, and the self-supervised contrastive CS-CAE, respectively.}
    \label{fig:distance}
\end{figure}

Figure~\ref{fig:distance} shows the detection efficiency as a function of source distance. Compared with a single global AUC value, the distance-dependent efficiency provides a more physically interpretable measure of sensitivity, because the observed CCSNe GW strain amplitude decreases with increasing distance. As expected, the detection efficiency decreases as the injected source distance increases. Under the adopted operating threshold, the characteristic distance corresponding to approximately $50\%$ detection efficiency is about $120\,\mathrm{kpc}$ for the proposed CS-CAE method.

At small distances, where the injected CCSNe signals have relatively high SNR, all three models recover a large fraction of the signals. The differences among the models become more informative near the transition region, where the signal amplitude becomes comparable to the background noise fluctuations. In this regime, CS-CAE maintains the highest detection efficiency over most of the tested distance range. This improvement suggests that the contrastive objective encourages the encoder to preserve waveform components that remain stable under independent noise realizations, thereby improving the effective detection range. Overall, the distance-dependent analysis shows that CS-CAE retains higher efficiency than both the supervised CNN and the conventional CAE over most of the tested distance range.

\begin{figure}[t]
    \centering
    \includegraphics[width=0.48\textwidth]{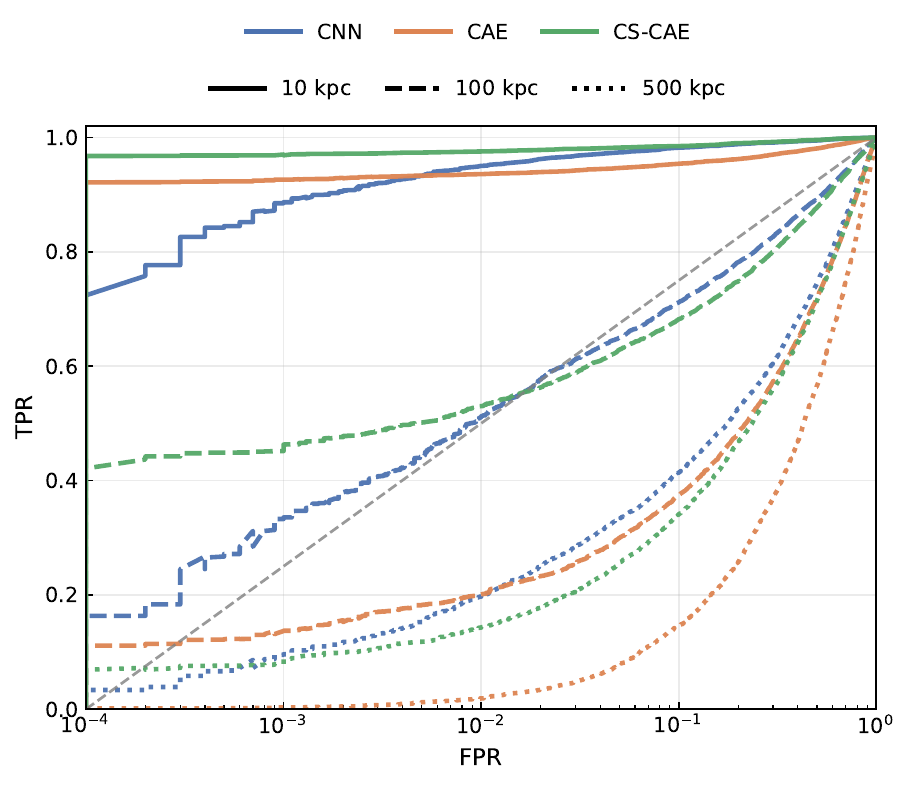}
    \caption{ROC curves at representative source distances of $10\,\mathrm{kpc}$, $100\,\mathrm{kpc}$, and $500\,\mathrm{kpc}$. Blue, orange, and green curves correspond to the CNN, CAE, and CS-CAE, respectively. Solid, dashed, and dotted line styles denote source distances of $10\,\mathrm{kpc}$, $100\,\mathrm{kpc}$, and $500\,\mathrm{kpc}$, respectively.}
    \label{fig:roc_distance}
\end{figure}

Figure~\ref{fig:roc_distance} further compares the ROC curves at representative source distances of $10\,\mathrm{kpc}$, $100\,\mathrm{kpc}$, and $500\,\mathrm{kpc}$. This comparison is particularly relevant for practical GW transient searches, because the low-false-positive-rate region of the ROC curve corresponds to the low-false-alarm regime required in observational analyses. A model with a high global AUC may still be suboptimal for real searches if its sensitivity degrades substantially after imposing a strict false-alarm constraint.

At $10\,\mathrm{kpc}$, the injected signals are strong, and all three models achieve high detection performance. At $100\,\mathrm{kpc}$, the separation among the models becomes more pronounced, with CS-CAE showing its clearest advantage in the low-FPR region. At $500\,\mathrm{kpc}$, which lies beyond the training-distance range, the overall true-positive rates are lower because of the weaker signals. Nevertheless, CS-CAE still retains a relative advantage in the low-FPR region. Therefore, the primary advantage of CS-CAE is not simply a marginal improvement in a global metric, but its ability to better preserve detection efficiency in the low-false-alarm regime where a realistic CCSNe GW search would operate.

\subsection{Detection statistics}
\label{subsec:statistics_representations}

\begin{figure*}[t]
    \centering
    \includegraphics[width=0.9\textwidth]{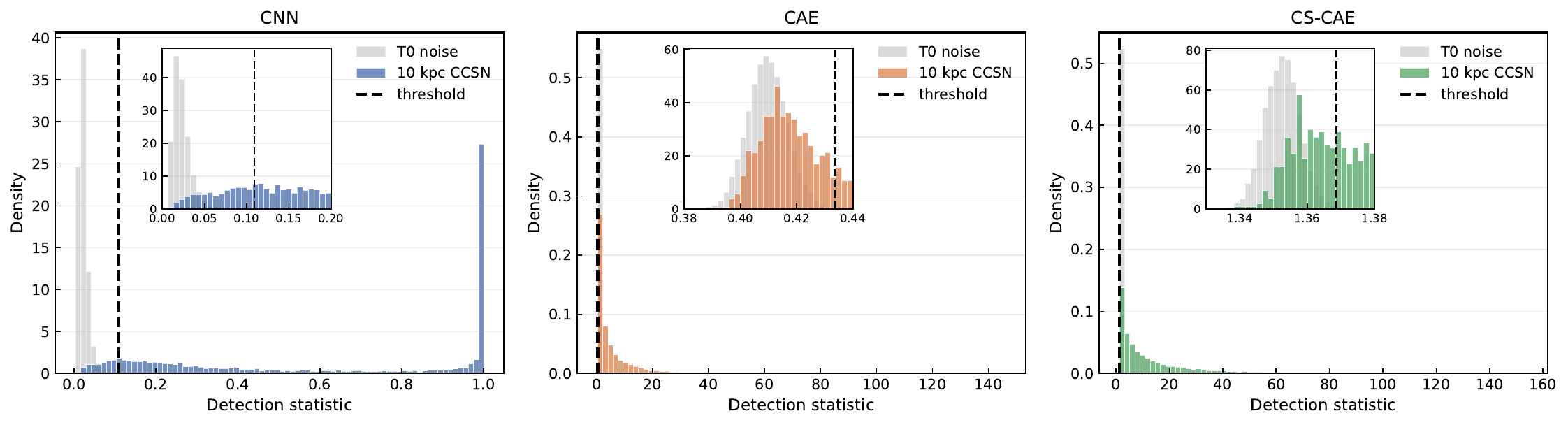}
    \caption{Detection-statistic distributions for the CNN, CAE, and CS-CAE. The left, middle, and right panels correspond to the CNN, CAE, and CS-CAE, respectively. Gray distributions denote pure-noise samples, colored distributions denote $10\,\mathrm{kpc}$ CCSNe samples, and black dashed vertical lines indicate the adopted decision thresholds. The inset in each panel shows an enlarged view of the threshold region.}
    \label{fig:metric}
\end{figure*}

Figure~\ref{fig:metric} compares the detection-statistic distributions for pure-noise samples and CCSNe injections. The detection statistics should be interpreted differently for the supervised and self-supervised models. For the supervised CNN, the statistic corresponds to a discriminative classification score optimized directly using binary labels, whereas for the CAE and CS-CAE, the statistics are derived from reconstruction behavior.

The pure-noise distributions are approximately Gaussian-like, consistent with the Gaussian background-noise setting adopted in the simulations. The CCSNe distributions are more sparsely populated toward larger statistic values, which is consistent with the log-uniform distance sampling of the test set and the resulting larger fraction of low-SNR injections. Compared with the conventional CAE baseline, the CS-CAE signal distribution is shifted more clearly toward larger statistic values, yielding better separation from the noise distribution.

\begin{figure*}[t]
    \centering
    \includegraphics[width=0.9\textwidth]{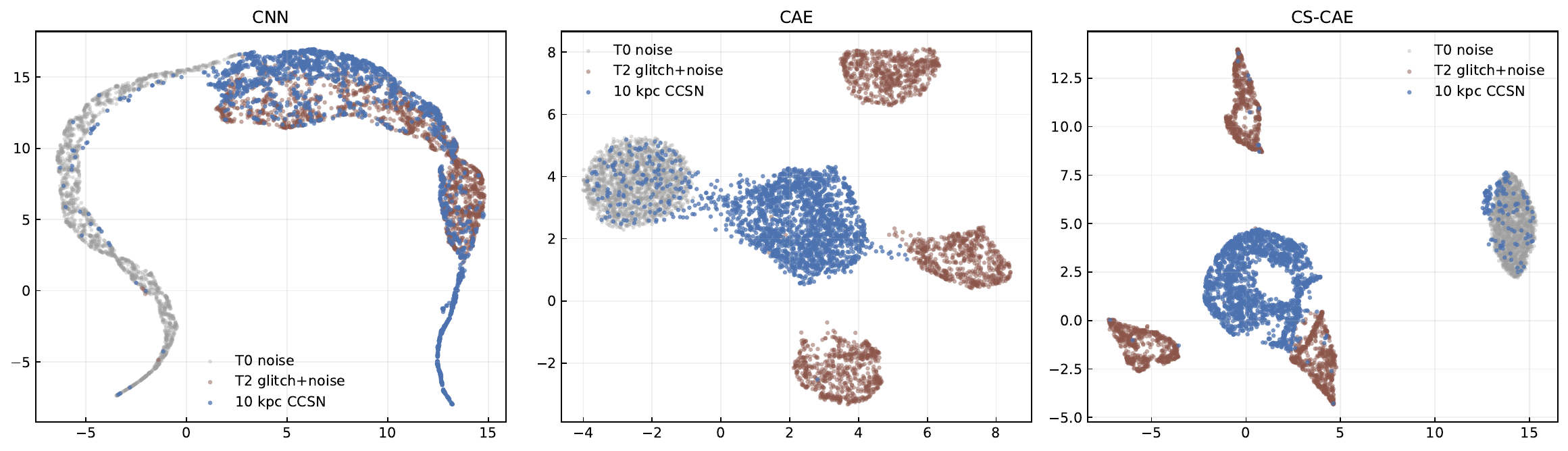}
    \caption{UMAP visualizations of learned representations for pure noise, glitch-plus-noise, and $10\,\mathrm{kpc}$ CCSNe samples. The left, middle, and right panels correspond to the CNN, CAE, and CS-CAE, respectively. Gray, brown, and blue points denote pure noise, glitch-plus-noise, and $10\,\mathrm{kpc}$ CCSNe samples, respectively.}
    \label{fig:umap}
\end{figure*}

The learned feature representations are visualized using uniform manifold approximation and projection (UMAP)~\cite{DBLP:journals/corr/abs-1802-03426} in Fig.~\ref{fig:umap}. The supervised CNN representation shows substantial overlap among pure noise, glitch-contaminated samples, and CCSNe signals, indicating that the label-trained classifier does not form a clean separation among these different inputs. The CAE representation improves the separation, but the CCSNe samples are still relatively extended and partially connected with background-like samples. In contrast, CS-CAE gives a clearer organization of the latent space. The CCSNe samples form a compact structure, while the glitch-contaminated samples are separated into several isolated clusters.

The separated glitch clusters have a direct physical interpretation. In the glitch-contaminated test set, a glitch is injected into one detector channel at a time. Therefore, the different glitch clusters in the UMAP space correspond to glitches appearing in different ET channels. This indicates that the learned representation does not treat glitches as coherent CCSNe-like signals. Instead, CS-CAE preserves information about the channel-localized nature of the transient. Thus, the UMAP result suggests that glitch candidates can be identified as channel-localized artifacts rather than multi-channel CCSNe-like events.

\begin{figure*}[t]
    \centering
    \includegraphics[width=0.9\textwidth]{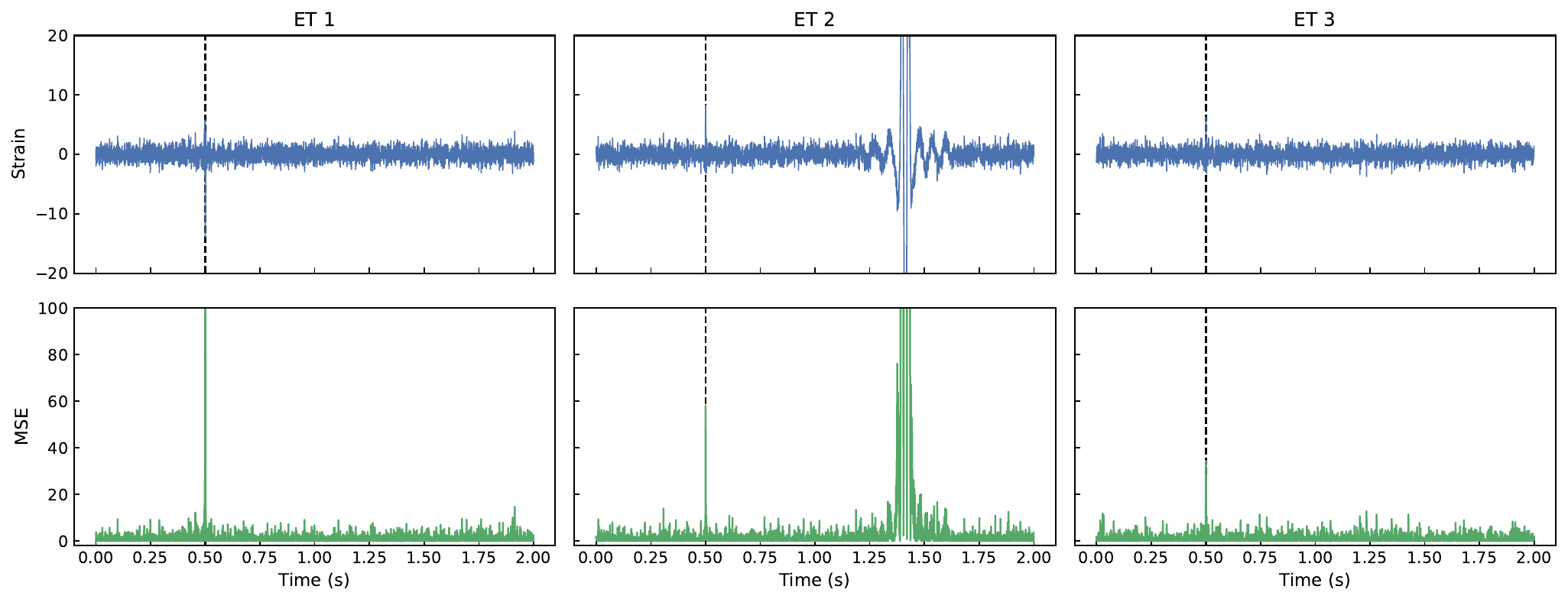}
    \caption{Example channel-wise CS-CAE response in the three ET channels. The left, middle, and right columns correspond to ET~1, ET~2, and ET~3, respectively. The upper panels show the whitened strain time series, and the lower panels show the corresponding mean-squared-error (MSE) reconstruction response. The glitch-like transient produces a localized excess mainly in one channel, while a CCSNe-like signal is expected to produce correlated responses across multiple channels. Black dashed vertical lines mark the reference transient times shown in each channel.}
    \label{fig:loss}
\end{figure*}

Figure~\ref{fig:loss} presents an example of the channel-wise CS-CAE response in the three ET channels. The upper panels show the whitened strain data, and the lower panels show the corresponding reconstruction-error response. This plot gives a more direct explanation of the glitch behavior observed in the UMAP space. The glitch-like transient produces a large reconstruction-error peak only in the affected ET channel, while the other channels do not show a consistent excess at the same time. Therefore, although the glitch is anomalous, it is a single-channel anomaly.

This behavior is different from the expected response of a genuine CCSNe GW signal. After projection onto the ET detector channels, a CCSNe signal should leave correlated responses across multiple channels. The channel-wise reconstruction-error pattern therefore provides a simple way to reject this type of glitch: an anomalous event dominated by only one channel can be removed as a glitch-like artifact, whereas an event with correlated responses in multiple ET channels is retained as a CCSNe-like candidate. This explains why the glitch samples can be excluded by channel-consistency information even though they may also produce large anomaly values.

\section{Conclusion}
\label{sec:conclusion}

In this work, we proposed a CS-CAE for the detection of CCSNe gravitational-wave signals in simulated ET data. The method combines a reconstruction-based detection statistic with a contrastive learning objective. By constructing positive pairs from independent noisy views of the same underlying CCSNe waveform, CS-CAE encourages the encoder to learn signal-consistent representations that are less sensitive to incidental detector-noise fluctuations.

The experimental results show that CS-CAE provides a favorable trade-off between in-distribution sensitivity and cross-waveform generalization. On the in-distribution simulated test set, the supervised CNN achieves the highest AUC, while CS-CAE remains close to the supervised CNN and clearly outperforms the conventional CAE. On the numerical CCSNe waveform test set, CS-CAE achieves the best AUC among the three models. We attribute this improved generalization to the noise-realization contrastive objective, which encourages the encoder to learn waveform features that are invariant to detector-noise fluctuations rather than features specific to the phenomenological simulation prescription used during training. Because the positive pairs are constructed from independent noisy realizations of the same signal without using waveform-family labels, the learned representations are less coupled to the particular waveform family seen during training and transfer more readily to numerical waveforms generated under different physical and numerical assumptions. This property is particularly important for CCSNe GW searches, where the true signal family is uncertain and cannot be fully represented by a single phenomenological simulation prescription.

The distance-dependent analysis further demonstrates that CS-CAE maintains higher detection efficiency over most of the tested distance range, with a characteristic $50\%$ detection-efficiency distance of approximately $120\,\mathrm{kpc}$ under the adopted operating threshold. The fixed-distance ROC curves show that the main advantage of CS-CAE appears in the low-FPR region, corresponding to the low-false-alarm regime required for practical GW transient searches. At the largest tested distance of $500\,\mathrm{kpc}$, although the absolute detection performance is reduced by the weaker signals, CS-CAE still preserves a relative advantage in the low-FPR region.

The detection-statistic distributions, UMAP visualizations, and channel-wise response examples provide additional insight into the learned behavior of the models. The supervised CNN output is a label-trained discriminative score, whereas the CAE and CS-CAE statistics are reconstruction-based and are therefore more naturally related to deviations from a learned noise reconstruction model. Among the three models, CS-CAE yields the clearest latent-space organization. In particular, the glitch-contaminated samples form separated clusters associated with glitches in different ET channels, while the CCSNe samples form a compact signal-like structure. The channel-wise reconstruction-error example further shows that the glitch response is mainly a single-channel anomaly. Therefore, such glitches can in principle be rejected by a channel-consistency check, while multi-channel correlated responses can be retained as CCSNe-like candidates.

The results demonstrate that contrastive self-supervised representation learning provides an effective strategy for improving the robustness and transferability of CCSNe GW detection models. By combining reconstruction-based anomaly sensitivity with contrastive constraints on signal-consistent latent representations, CS-CAE achieves competitive in-distribution performance, improved generalization to numerical CCSNe waveforms, enhanced sensitivity in the low-FPR regime, and clearer diagnostic separation between multi-channel CCSNe-like responses and single-channel glitch-like artifacts. These properties are particularly valuable for CCSNe searches, where the signal morphology is uncertain, the available waveform catalogs remain incomplete, and fully supervised models may overfit to the adopted simulation prescription. Several limitations nevertheless remain. To maintain a controlled architectural comparison with the supervised CNN and the CAE baseline, the self-supervised models in this work are also implemented with convolutional networks. Future work will extend the present framework to long short-term memory-based self-supervised architectures, motivated by their strong performance in related sequence-modeling and gravitational-wave anomaly-detection studies~\cite{Moreno:2021fvp,Raikman:2023ktu,Iess:2023quq}. The present CS-CAE processes the three ET channels jointly and does not explicitly enforce cross-channel coherence, which may limit its rejection of glitch morphologies that partially resemble weak CCSNe signals. Future work will incorporate detector-wise feature extraction, cross-channel coherence constraints, and glitch-aware contrastive training to further improve robustness against instrumental transients. Taken together, these results highlight the potential of contrastive self-supervised learning to enable more robust, transferable, and less template-dependent CCSNe gravitational-wave searches with next-generation detectors.

\section*{Acknowledgements}
This work was supported by the National Natural Science Foundation of China (Grants Nos. 12473001, 12575049, and 12533001), the National SKA Program of China (Grants Nos. 2022SKA0110200 and 2022SKA0110203), the China Manned Space Program (Grant No. CMS-CSST-2025-A02), and the 111 Project (Grant No. B16009).

\section*{Data Availability}
The data supporting the results of this article will be made publicly available in an online repository upon publication.

\bibliography{ccsn_find}

\end{document}